\documentclass[onecolumn,epsfig,amsmath,showpacs,amssymb]{revtex4}

\usepackage{epsfig}

\begin{document}

\title{Anomalous nucleation far from equilibrium}

\author{I.~T.~Georgiev, B.~Schmittmann, and R.~K.~P.~Zia}
\affiliation{Center for Stochastic Processes in Science and Engineering and\\
Department of Physics, Virginia Tech, Blacksburg, VA 24061-0435 USA}
\date{\today}

\begin{abstract}

We present precision Monte Carlo data and analytic arguments for an 
asymmetric exclusion process, involving two species of particles driven
in opposite directions on a $2 \times L$ lattice. 
We propose a scenario which resolves a stark discrepancy between 
earlier simulation data, suggesting the existence of an ordered phase, 
and an analytic conjecture according to which the system should revert to a
disordered state in the thermodynamic limit. 
By analyzing the finite size effects in detail, we argue that the presence
of a single, seemingly macroscopic, cluster is an intermediate stage of a 
complex nucleation process: In smaller systems, this cluster is 
destabilized while larger systems allow the formation of multiple clusters. 
Both limits lead to exponential cluster size distributions which are, however, 
controlled by very different length scales. 

\end{abstract}

\vspace{1mm} 

\maketitle
\vspace{-1mm}

{\it Introduction.} From recent studies of statistical model systems far
from thermal equilibrium, it is clear that even their stationary states pose
many new challenges. The well established theoretical
machinery for systems in thermal equilibrium is essentially powerless in
these contexts, and our equilibrium-trained intuition often misleads us. For
example, though the specific heat of an equilibrium system cannot be
negative, the internal energy of a non-equilibrium steady state may decrease
when coupled to a thermal bath at a higher temperature \cite{AJP}. In
general, non-equilibrium steady states depend sensitively on the details of
the microscopic dynamics, resulting in widely diverse behavior at the
macroscopic level. So far, an overarching framework is still lacking, 
but much insight has been gained by studying simple model systems.

One such class of models are driven diffusive systems \cite{KLS}.
Close relatives of the asymmetric exclusion
process \cite{ASEP}, they involve interacting diffusing particles, driven
into selected spatial directions by an external force. 
Models of this kind have been invoked to describe vehicular and 
pedestrian traffic \cite{traffic}, gel electrophoresis \cite{gel}, 
and a wide range of biological problems, ranging from molecular 
motors \cite{motors} to protein synthesis \cite{proteins}.
A particularly rich
phase diagram emerges for two species of particles, driven in opposite
directions on a two-dimensional ($2D$) periodic lattice \cite{SHZ}. Even
with no interaction except excluded volume, we observe
transitions from a freely flowing homogeneous state to a ``jammed'' 
state displaying a {\em macroscopic} cluster of particles. 
These transitions persist even if neighboring particles 
are allowed to exchange places, with a small rate $\gamma $
\cite{KSZ1}. In contrast, an analytic solution \cite{SG} for the
same system in $1D$ shows that the homogeneous state prevails always, for
any $\gamma >0$ and particle density $\rho <1$. Exploring notions of a lower 
critical dimension, we studied a {\em quasi-}$1D$ system, involving just two
``lanes'', i.e., $2\times L$, with $L$ up to $10^{4}$ \cite{KSZ2,Jay}.
Remarkably, for small $\gamma $ and half-filling, the {\em jammed}
state re-emerged, with the length of the jam scaling with $L$. However, it
was argued recently \cite{Kafri} that this jam is a mere finite-size effect
and should break up into a disordered state when $L$ exceeds a characteristic
crossover length $L_{c}$. While the latter is not known exactly, it might 
be as large as $10^{70}$ \cite{Rajewsky}!

Since simulations are an essential tool in the study of nonequilibrium
steady states and should provide a fairly accurate picture of the
thermodynamic limit, such stark discrepancies between simulation data and
analytic arguments are disconcerting and need to be explored further. In
this letter, we report {\em extensive, high precision }computer simulations,
studying the full $(\rho ,\gamma )$ space and a wide range of $L$ for this 
two-lane system. Defining an ``$s$-cluster'' as $s$
particles of either species, connected to each other via nearest-neighbor
bonds, we focus on the {\em normalized} residence distribution, $p(s)$,
which is the probability that a randomly chosen particle belongs to an $s$%
-cluster. For a homogeneous state, $p(s)$ is monotonically decreasing. In
contrast, for a jammed system it displays {\em two} distinct peaks: one at
the origin ($s=1$) and another marking the macroscopic cluster, at 
$s_{o}\propto L$ \cite{KSZ2}. If a crossover length $L_{c}$ exists, this 
second peak should disappear when $L$ increases beyond $L_{c}$. Expecting 
$L_{c}$ to decrease with $\gamma $, we consider much larger $\gamma $'s 
than in previous studies \cite{KSZ2} and find, indeed, that systems with 
$\rho > 0.3 $ cross over to monotonic distributions with
increasing $L$. However, \ at {\em lower} densities (e.g., $\rho =0.1$) we
discover the {\em opposite} behavior: here, $p(s)$ is monotonic at small $L$
and develops a second peak as $L$ becomes larger. In order to interpret
these complex finite-size effects, we propose a new picture, in which larger
clusters compete with a homogeneous background of ``travellers'', consisting
only of isolated particles and very small clusters. 
In the remainder of this letter, we briefly review the model, present
the simulation results, and provide the analysis that leads us to our
conclusions.

\begin{figure*}[!t]
\epsfig{file=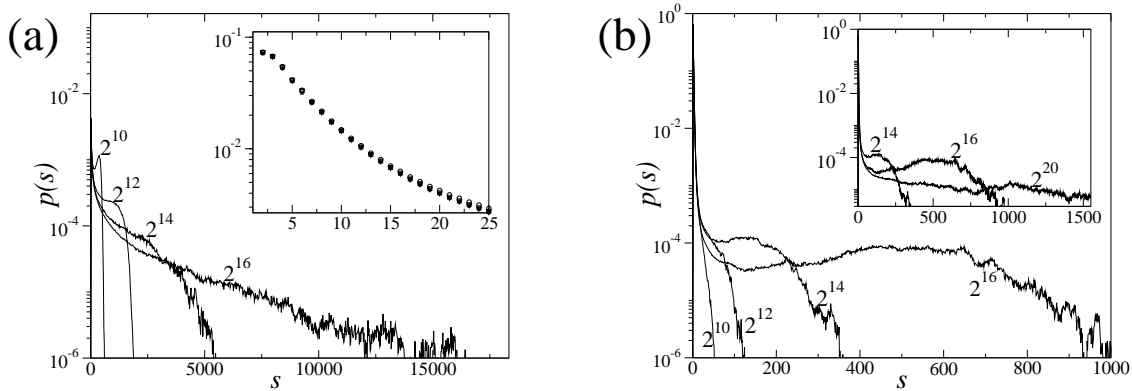, width=6in}
\caption{The residence distribution $p(s)$ vs.~$s$ on lattices with $L=2^{10}$, $2^{12}$,
$2^{14}$ and $2^{16}$ for (a) $\rho=0.5$ and $\gamma=0.45$ and (b) $\rho=0.1$ and $\gamma=0.163$.
The inset on the left figure magnifies the small-$s$ data and the inset on the right
shows the change from a double-peaked to a single-peaked distribution by increasing
$L$.}
\label{fig1}
\end{figure*}

{\it The model.} In a $2\times L$ lattice with fully periodic boundary 
conditions, each site can be either empty ($\oslash $) or occupied
by a single particle, which we identify as either ``positive'' ($+$) or
``negative'' ($-$). Configurations are labeled by 
$\left\{\sigma _{x,y}\right\} $, 
where $\sigma $ takes on values $0,\pm 1$ and 
$x\in \left[ 1,L\right] $; $y=1,2$. 
The two species are ``driven'' in opposite
directions, much like cars on a 2-lane road. Starting from an
initial configuration with {\em equal} numbers of the two species occupying a
fraction, $\rho $, of the sites (i.e., $N_{+}=N_{-}=\rho L$), we evolve the
system by exceedingly simple rules. In one Monte Carlo step (MCS), the
following steps are repeated $2L$ times: ({\em i}) a pair of two
nearest-neighbor sites (a ``bond'') is chosen at random; ({\em ii}) for
bonds along the $L$ direction, a ($+\oslash $) or a ($\oslash -$) pair is
always exchanged while a ($+-$) pair is exchanged with probability $\gamma$;
({\em iii}) for transverse bonds, particle-hole pairs are always
exchanged while particle-particle pairs are exchanged with probability 
$\gamma $. Other moves are not allowed.

To probe the system's behavior and the effects of finite size, we explore a
wide range of $\gamma $ and $\rho $, and lattices with $L=2^{8}$, $2^{10}$, 
$2^{12}$, $2^{14}$, and $2^{16}\simeq 64K$. Since our main focus is the
steady state, we discard up to $2\times 10^6$ MCS before taking measurements.
Thereafter, quantities of interest are recorded every $100$ MCS, for up
to $10^7$ MCS. On a $64$-bit machine, we exploit a fast multispin coding
algorithm (e.g., \cite{newman}) by encoding the state on a site for $64$
different lattices (realizations) simultaneously. Thus, we are confident
that our statistical errors are minimal. 

{\it Simulation results.} We set the scene by presenting a characteristic
data set for the residence distribution $p(s)$ at $\rho =0.5$ and $\gamma
=0.45$ (Fig.\ \ref{fig1}a), in order to demonstrate the crossover from bimodal to
single-peaked distributions with {\em increasing} $L$. The smallest system,
$L=2^{10}$, shows a distinct peak at $s_{o}\simeq 380$, while the next size,
$L=2^{12}$, shows only a shoulder which broadens into a well-defined
exponential, $\propto \exp \left[ -s\xi (\gamma ,\rho )\right] $, with a
`slope' $\xi (0.45,0.5)\simeq 0.25 \times 10^{-3}$ for the largest system.
It is remarkable that, even for such large $L$'s, the large $s$ data
still depend quite sensitively on $L$. On the other hand, as the inset shows, 
the small $s$ part is manifestly independent of $L$, a feature that persists
in the whole ($\gamma ,\rho $) space.
Since we expect the crossover length $L_{c}$ to decrease with $\gamma$, 
we consider systems with $\gamma =0.5$, keeping $\rho$ fixed at $0.5$.
Here, even the smallest system ($L=2^{10}$) already displays a monotonic
distribution. In contrast, if $\gamma =0.4$, the distributions remain
double-peaked for all $L \leq 2^{16}$. In fact, for those $L$,  
the (second) peak position scales as $s_{o}\sim 0.8\,L^{0.95}$ which
one might (naively) interpret as the signature of a ``macroscopic'' cluster,
unless even larger $L$ are investigated. 
To summarize, these findings are consistent with the conjecture 
\cite{Kafri} that the presence of 'macroscopic' clusters is a finite-size effect. 
For $L\rightarrow \infty $, we may expect residence distributions to comprise two
components: one at small $s$ (which we associate with the ``travellers'') and 
a slow exponential decay $\exp \left[ -s\xi (\gamma ,\rho )\right] $ 
(associated with large scale ``jams'').

Surprisingly, the behavior of our model is richer than this simple picture 
suggests. For smaller density and $\gamma$, there is an {\em additional} 
crossover, in the {\em opposite} direction: Distributions are monotonic in small 
$L$ systems and become bimodal as $L$ {\em increases}, only reverting to being 
monotonic for much larger $L$. The first of these crossovers is illustrated 
in Fig.\ \ref{fig1}b, which shows residence distributions for the same four 
$L$'s but at $\rho =0.1$ and $\gamma =0.163$. We re-emphasize: In contrast 
to Fig.\ \ref{fig1}a, here the second peak emerges with {\em increasing} $L$.

\begin{figure}[!t]
\epsfig{file=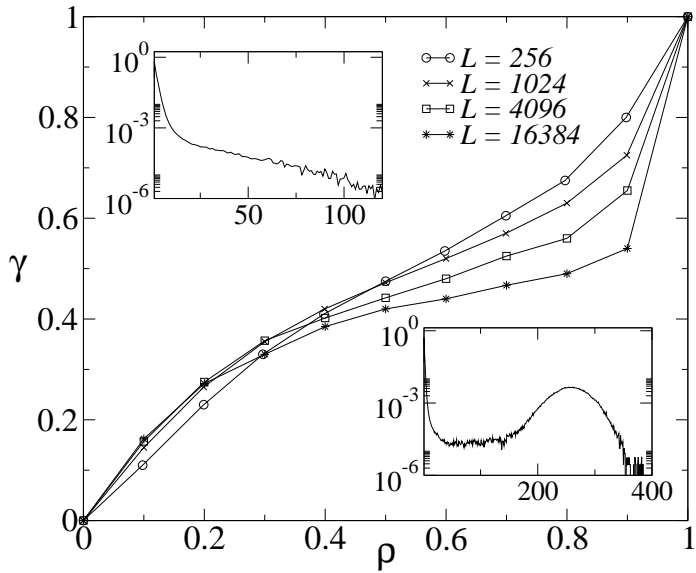,angle=-90, width=4in}
\caption{Boundaries separating monotonic from bimodal distributions, for 
various $L$. The insets show $p(s)$ for $\rho=0.1$, $\gamma=0.3$ (top left) and $\rho=0.1$,
$\gamma=0.125$ (bottom right) on a $L=2^{12}$ lattice.}
\label{fig2}
\end{figure}

Fig.\ \ref{fig1} suggests that, for any given $L$, it is possible to identify two
distinct regions in ($\gamma ,\rho $) space: In one of them, located near
the large $\gamma $, small $\rho $ corner, $p\left( s\right) $ is
monotonically decreasing in $s$; in the other region (small $\gamma $,
larger $\rho $), $p\left( s\right) $ is bimodal. To summarize our data, the
corresponding boundaries, $\gamma_{c}\left(\rho,L \right)$, 
are drawn in Fig.\ \ref{fig2} for several $L$. If larger
systems favor (disfavor) bimodal distributions, $\gamma_{c}\left(\rho,L \right)$ 
shifts up (down) with $L$.

\begin{figure}[!t]
\epsfig{file=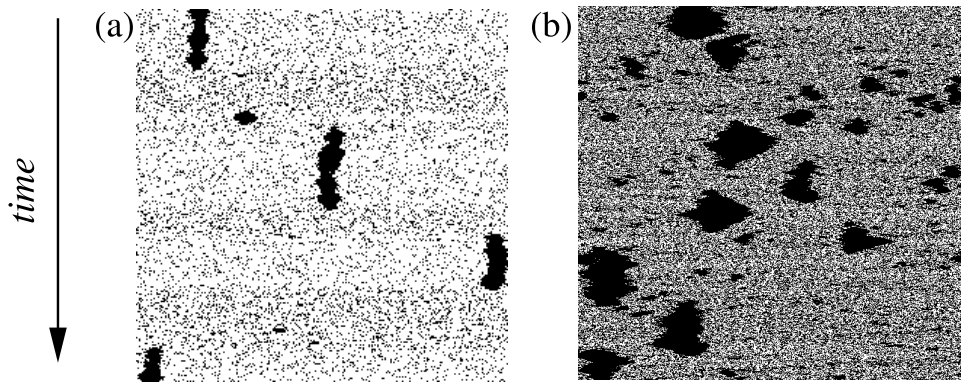,width=5in}
\caption{Space-time plot at $\rho =0.1$, $\gamma =0.1$, $L=2^8$ (a) and $\rho =0.4$, $\gamma
=0.4$, and $L=2^9$ (b). The spatial (temporal) coordinate runs horizontally (vertically, in
units of $50$ MCS). A black dot indicates a particle in either lane.}
\label{fig3}
\end{figure}

{\it Analytic arguments. }In the following, we present a simple scenario
which encompasses the full range of these phenomena. Our discussion begins
at very low densities. A space-time plot (Fig.\ \ref{fig3}a) for $\rho =0.1$, $\gamma
=0.1$, and $L=2^8$ shows the evolution of a typical configuration. 
With most of the system occupied by travellers, local density
fluctuations occasionally ``nucleate'' a single larger cluster, containing
about $24\pm 8$ particles, which survives for up to $1000$
MCS, then disintegrates and reforms elsewhere. When one of these large
clusters is present, the background density of travellers is depleted
noticeably. Reducing $\gamma $ lengthens the cluster life times,
until a single cluster persists for the whole run. In contrast, at higher $%
\gamma $, large clusters just blink on and off, and the traveller 
component dominates.

Motivated by these observations, we describe the system at such low
densities in terms of two ``components''. One of these is
purely uniform, consisting only of travellers at density $\rho $. The other
exhibits a single large cluster, of mass $s$, reducing the density of
travellers in the remainder of the system to $\rho (s)=\left(
2L\rho -s\right) /\left( 2L-s\right) $. In the spirit of mean-field theory,
we neglect fluctuations in the cluster size as well as any charge
asymmetries. Since the total number of each species is conserved, it is
natural to focus on, e.g., the current of positive particles through either
lane, as a function of the local densities. Of course, the current in the
interior of a cluster, $j_{c}(s)$, should display a very different density
dependence from the current carried by the travellers, $j_{t}(s)$. Assuming
the traveller region to be approximately uniform, $j_{t}(s)$ should be well
represented by its mean-field form \cite{KSZ1}, $j_{t}(s)=\rho (s)\left[
1-\rho (s)\right] /2+(\gamma /4)\rho ^{2}(s)$. Here, the first term reflects
particle-hole exchanges, proportional to the density of positive particles, $%
\rho (s)/2$, and that of holes, $1-\rho (s)$. The second term models the
charge exchanges. For the current through a cluster of size $s$, the exact
form is known only for the one-lane model, $j_{c}(s)=(\gamma /4)\left[
1+b/s+O(s^{-2})\right] $ with $b=3/2$ \cite{Kafri}. For the
two-lane system, we have to rely on simulations which confirm the same form
up to a slightly modified coefficient, $b\simeq 1.6$ \cite{Kafri, current}.
Even though the $O(s^{-2})$ corrections are not taken into account, this
form appears to be sufficiently accurate for our purposes.

Clearly, a large cluster of size $s$ can co-exist with a uniform region only
if the two currents, $j_{c}(s)$ and $j_{t}(s)$, are equal. If $%
j_{c}(s)>j_{t}(s)$, the cluster will lose particles and shrink; for $%
j_{c}(s)<j_{t}(s)$, it will grow in size. This allows us to identify both 
{\em stable} and {\em unstable }solutions of $j_{c}(s)=j_{t}(s)$, as
functions of $\gamma $, $\rho $, and $L$, in the physical domain $s\geq 1$.
Leaving details to a future publication \cite{GSZ}, we just outline the main
qualitative predictions here. To begin with, we note that $j_{c}(s)$ is
independent of $\rho $ and $L$ while $j_{t}(s)$ varies only very weakly with 
$\gamma $. In other words, we can plot $j_{t}(s)$ for a given ($\rho ,L$)
and then watch $j_{c}(s)$ shift up or down, proportional to $\gamma $. In
this fashion, we can easily identify three regimes: ({\em i}) At small $\rho 
$ and $L$, and $\gamma $ sufficiently large, no solution exists: $j_{t}(s)$
remains well below $j_{c}(s)$, and the system is homogeneously filled with
travellers. While local fluctuations occasionally generate small clusters,
these disintegrate almost immediately. The two smallest system sizes of
Fig.\ \ref{fig1}b belong into this regime. ({\em ii}) As $\gamma $ decreases, the two
curves $j_{t}(s)$ and $j_{c}(s)$ first touch, and then two solutions emerge:
an unstable one ($s_{u}$) close to the origin which plays the role of a
critical droplet, and a stable one at $s_{o}>s_{u}$ which sets the cluster
size. If $s_{o}$ is not much larger than $s_{u}$ and fluctuations are
included, we expect to see purely uniform configurations compete with those
supporting one larger cluster, as in Fig.\ \ref{fig3}a. For those parameter values, we
predict $s_{o}=24$, in remarkably good agreement with the data. 
Returning to Fig.\ \ref{fig1}b, we note that two solutions can also
emerge if $L$ {\em increases} at fixed $\gamma $, as illustrated by the $L=2^{12}$
data. Analytically, we expect $s_{o}=70$ here. Since the two
currents differ by less than $5$ up to $s\simeq 120$, clusters of all
these sizes form quite easily, giving rise to the observed $p(s)$. ({\em iii}%
) Finally, as $\gamma $ is lowered further, only the stable solution
survives and shifts to $s_{o} \sim O(L)$. Before this occurs,
however, fluctuations begin to play a much more important role: Considering
the $L=2^{16}$ system in Fig.\ \ref{fig1}b, our theory would predict $s_{o}=1436$, but the
largest {\em observed} clusters remain well below $700$ particles.  
Monitoring typical configurations, we discover that the system is attempting 
to nucleate a {\em second} cluster in the traveller region, in addition to a 
persistent first one. 

To summarize, our simple mean-field theory describes the ``nucleation'' of a
single cluster, from a uniform background of travellers, remarkably well.
Naturally, once a single cluster exists, and either $\rho $ or $L$ are
increased (or $\gamma $ reduced), we should expect, and do indeed observe, 
the ``nucleation'' of additional clusters in the traveller region. As this
process continues, the residence
distribution broadens, and eventually crosses over to a slow exponential
decay, as in Fig.\ \ref{fig1}a. To illustrate this, we present a space-time plot for 
$\rho =0.4$, $\gamma =0.4$ and $L=2^9$ (Fig.\ \ref{fig3}b) which shows several 
clusters of different sizes in each time slice.

{\it Conclusions.} We have presented extensive simulation data for a $%
2\times L$ system with two species of particles driven in opposite
directions. Varying $L$, density $\rho $, and the particle exchange
rate $\gamma $, we monitor the cluster size
(``residence'')\ distribution $p(s)$. We find two well-separated domains in (%
$\rho ,\gamma $) space: one at low $\rho $, high $\gamma $ with
monotonically decreasing $p(s)$, and one at high $\rho $, low $\gamma $ with
bimodal distributions. The boundary between these domains shifts with $L$,
the general trend favoring monotonic $p(s)$ as $L$ increases. 

To understand the underlying physics at the crossovers, we draw some 
insight from nucleation processes. For any fixed value of $\gamma $,
systems are uniform for sufficiently low $\rho $ and $L$, and the associated
residence distribution falls off rapidly. As either $\rho $ or $L$
increases, a single large cluster is nucleated. The residence distribution
develops a distinct shoulder and, eventually, a second peak. This transition
is well captured by a simple mean-field theory which balances the current
through the cluster with that through the uniform region. As $\rho $ or $L$
increase further, a second cluster is nucleated, and eventually a whole
population is established. As a result, the residence distribution becomes
monotonic again, exhibiting a {\em slow} exponential decay. 

With presently accessible $L$'s, we may only conclude that 
our data for $ \gamma_{c} \left( \rho,L \right)$ (cf.~Fig.\ \ref{fig3})
is {\em consistent} with the conjecture \cite{Kafri}. 
At the same time, given the absence of an
exact solution, we cannot rule out the possibility that 
$\gamma_{c} \left( \rho,L \right)$ might 
have a {\em non-trivial} $L \rightarrow \infty$ limit. 
We do stress, however, that its {\em non-monotonic} 
$L$-dependence (at fixed $\rho$) appears unprecedented in equilibrium
nucleation \cite{PAR} or phase separation \cite {CSN} processes. 
Whatever the eventual conclusion, we believe that
it is important to explore finite-size effects, since these
are relevant for most applications of interest. For example,
in the modelling of protein synthesis, system sizes ranging from $1K$ to $10M$ 
cover the typical lengths of mRNA molecules.
In these contexts, we feel that both crossovers may be of theoretical as
well as practical interest. 

\emph{Acknowledgements.} We thank J.T. Mettetal and M. Mobilia 
for stimulating discussions. This research is supported in part 
by the US National Science Foundation 
through grants DMR-0088451 and DMR-0414122.


\begin{thebibliography}{99}

\bibitem{AJP} R.K.P. Zia, E.L. Praestgaard, and O.G. Mouritsen, 
Am. J. Phys. {\bf 70}, 384 (2002). 
\bibitem{KLS} S. Katz, J.L. Lebowitz, and H. Spohn, Phys. Rev. B 
{\bf 28}, 1655 (1983) and J. Stat. Phys. {\bf 34}, 497 (1984).
For reviews, see, e.g., B. Schmittmann and R.K.P. Zia, 
Statistical mechanics of driven diffusive systems. 
In: {\em Phase Transitions and Critical Phenomena}, 
Vol. 17, ed. by C. Domb and J.L. Lebowitz (Academic Press, New York 1995); 
D. Mukamel, Phase transitions in non-equilibrium systems,
In: {\em Soft and Fragile Matter: Nonequilibrium Dynamics, Metastability
and Flow}, eds. M.E. Cates and M.R. Evans (IOP Publishing,
Bristol, 2000).

\bibitem{ASEP} J.T. MacDonald, J.H. Gibbs, and A.C. Pipkin, Biopolymers 
{\bf 6}, 1 (1968);  
for a review, see G.M. Sch\"{u}tz, Integrable stochastic
many-body systems. In: {\em Phase Transitions and Critical Phenomena}, Vol.
19, ed. by C. Domb and J.L. Lebowitz (Academic Press, New York 2001).

\bibitem{traffic} O. Biham, A.A. Middleton, and D. Levine: Phys. Rev. A
{\bf 46}, R6124 (1992); 
D. Chowdhury, L. Santen, and A. Schadschneider: Phys. Rep. 
{\bf 199} (2000).

\bibitem{gel} M. Rubinstein: Phys. Rev. Lett. \textbf{59}, 1946 (1987);
T.A.J. Duke: Phys. Rev. Lett. {\bf 62}, 2877 (1989); 
B. Widom, J.L. Viovy, and A.D. Desfontaines: J. Phys I
(France) {\bf 1}, 1759 (1991); U. Alon and D. Mukamel Phys. Rev. E 
{\bf 55}, 1783 (1997).

\bibitem{motors} J. Howard, Nature {\bf 389}, 561 (1997); M.E. Fisher and 
A.B. Kolomeisky, Proc. Natl. Acad. Sci. USA {\bf 96}, 6597 (1999).

\bibitem{proteins} L.B. Shaw, R.K.P. Zia, and K.H. Lee, 
Phys. Rev. E {\bf 68}, 021910 (2003); 
MacDonald {\em et.al.} in \cite{ASEP}.

\bibitem{SHZ} B. Schmittmann, K. Hwang and R.K.P. Zia, 
Europhys. Lett. {\bf 19}, 19 (1992).

\bibitem{KSZ1} G. Korniss, B. Schmittmann and R.K.P. Zia, J. Stat. Phys.
{\bf 86}, 721 (1997).

\bibitem{SG} S. Sandow and C. Godr\`{e}che, unpublished (1998).

\bibitem{KSZ2} G. Korniss, B. Schmittmann and R.K.P. Zia, Europhys.
Lett. {\bf 45}, 431 (1999).

\bibitem{Jay} J. T. Mettetal, B. Schmittmann and R.K.P. Zia, 
Europhys. Lett. {\bf 58}, 653 (2002); B. Schmittmann, J.T. Mettetal and 
R.K.P. Zia, in {\em Computer Simulation Studies in Condensed Matter Physics
XVI}, eds. D.P. Landau, S.P. Lewis, and H.B. Sch{\"{u}}ttler (Springer, in
press).

\bibitem{Kafri} Y. Kafri, E. Levine, D. Mukamel, G.M. Sch{\"{u}}tz and 
J. T{\"{o}}r{\"{o}}k, Phys. Rev. Lett. {\bf 89}, 035702 (2002); Y. Kafri, E.
Levine, D. Mukamel and J. T{\"{o}}r{\"{o}}k, J. Phys. A: Math. Gen. 
{\bf 35}, L459 (2002).

\bibitem{Rajewsky} N. Rajewsky, T. Sasamoto and E.R. Speer, Physica A 
{\bf 279}, 123 (2000); T. Sasamoto and D. Zagier, J. Phys. A: Math. Gen. 
{\bf 34}, 5033 (2001).

\bibitem{newman} M.E.J. Newman and G.T. Barkema, {\em Monte Carlo Methods
in Statistical Physics} (OUP, 1999).

\bibitem{current} J.T. Mettetal, B. Schmittmann and R.K.P. Zia, unpublished;
M.R. Evans, E. Levine, P.K. Mohanty, and D. Mukamel, cond-mat/0405049.

\bibitem{GSZ} I. Georgiev, B. Schmittmann and R.K.P. Zia, in preparation.

\bibitem{PAR} P.A. Rikvold, H. Tomita, S. Miyashita and S.W. Sides, 
Phys. Rev. E {\bf 49}, 5080 (1994).

\bibitem{CSN} A. Christensen, P. Stoltze and J.K. Norskov, 
J. Phys.: Cond. Mat. {\bf 7}, 1047 (1995).

\end{thebibliography}
\end{document}